\documentclass[aps,prd,preprint,a4paper,showpacs,superscriptaddress,nofootinbib,floatfix]{revtex4}
\usepackage{latexsym}
\usepackage{graphics}
\usepackage{graphicx}
\usepackage{bm}
\usepackage{amsmath}
\usepackage{amsfonts}
\usepackage{amssymb}

\begin{document}
\newcommand{\missE}{E \hspace{-0.65em}/}
\newcommand{\pslash}{p\hspace{-2mm}/ \hspace{0mm}}
\newcommand{\be}{\begin{eqnarray}}
\newcommand{\ee}{\end{eqnarray}}
\newcommand{\nn}{\nonumber}
\newcommand{\ovl}{\overline}
\newcommand{\ra}{\rightarrow}
\newcommand{\lra}{\longleftrightarrow}
\newcommand{\ba}[1]{\begin{eqnarray} \label{(#1)}}
\newcommand{\ea}{\end{eqnarray}}
\newcommand{\rf}[1]{(\ref{(#1)})}
\newcommand{\dps}{\displaystyle}
\def\mNl{m_{\tilde{\nu}_\ell^N}}
\def\mM{\tilde{m}_M}
\def\0n{0\nu\beta\beta}
\def\LM{$L\hskip-2mm /\mbox{MSSM}\;$}
\def\mN{m_{\tilde{\nu}_\ell^N}}
\def\mD{\tilde{m}_D}
\def\Lbar{L\hskip-1.5mm/}
\def \lsim {\mbox{${}^< \hspace*{-7pt} _\sim$}}
\def \gsim {\mbox{${}^> \hspace*{-7pt} _\sim$}}
\def \leql { ^< \hspace*{-7pt} _=}
\def \geql { ^> \hspace*{-7pt} _=}
\def\rp{$R_p \hspace{-1em}/\;\:$}
\def\rpm{R_p \hspace{-0.8em}/\;\:}
\def\rpt{$R_p \hspace{-0.85em}/\ \ $}
\def\et{E_T \hspace{-1em}/}
\def\d{\partial \hspace{-0.55em}/}
\def\pmb#1{\setbox0=\hbox{#1}%
\kern-.015em\copy0\kern-\wd0
\kern.03em\copy0\kern-\wd0
\kern-.015em\raise.0233em\box0 }
\def\olrap{\overleftrightarrow{\partial}}
\def \znbb {0\nu\beta\beta}
\def \tnbb {2\nu\beta\beta}
\def \emass {\langle m_{\nu} \rangle}
\def\bfr{\pmb{${r}$}}
\def\bfsgm{\pmb{${\sigma}$}}
\def\sir{({ \bfsgm_{i}^{~}} \cdot {\hat{\bfr}_{ij}^{~}} )}
\def\sjr{({ \bfsgm_{j}^{~}} \cdot {\hat{\bfr}_{ij}^{~}} )}
\def\si{{ \bfsgm_{i}^{~}}}
\def\sj{{ \bfsgm_{j}^{~}} }
\hyphenation{re-la-ti-vis-tic}
\hyphenation{struc-ture}
\hyphenation{char-gi-nos}
\hyphenation{vio-la-ting}
\hyphenation{to-po-lo-gy}
\hyphenation{ava-i-lable}

\title{
Testing SUSY models of lepton flavor violation at a photon collider
}

\author{M.~Cannoni}
\author{C.~Carimalo}
\author{W.~Da~Silva}
\affiliation{ Laboratoire de Physique Nucl\'eaire et de Hautes Energies,\\
IN2P3 - CNRS, Universit\'e Paris VI et VII, 4 Place Jussieu,
75525 Paris cedex 05, France}
\author{O.~Panella}
\affiliation{Istituto Nazionale di Fisica Nucleare, Sezione di
Perugia, Via A.~Pascoli, I-06123, Perugia, Italy}

\begin{abstract}
{The loop level lepton flavor violating signals $\gamma \gamma \to
\ell \ell' \;(\ell=e,\mu,\tau, \ell \neq \ell^\prime)$ are studied
in a scenario of low-energy, R-parity conserving, supersymmetric
see-saw mechanism within the context of a high energy photon
collider. Lepton flavor violation is due to off diagonal elements in
the left s-lepton mass matrix induced by renormalization group
equations. The average slepton masses ${\widetilde{m}}$ and the off
diagonal matrix elements $\Delta m$ are treated as model independent
free phenomenological parameters in order to discover regions in the
parameter space where the signal cross section may be observable. At
the energies of the $\gamma \gamma$ option of the future high-energy
linear collider the signal has a potentially large standard model
background, and therefore particular attention is paid to the study
of kinematical cuts in order to reduce the latter at an acceptable
level. We find, for the ($e\tau$) channel, non-negligible fractions
of the parameter space ($\delta_{LL}=\Delta m^2/\widetilde{m}^2
\gtrsim 10^{-1}$) where the statistical significance ($SS$) is $SS
\gtrsim 3$.}
\end{abstract}


\maketitle

\section{Introduction}
\label{intro}

The high-energy linear lepton collider (LC) is presently considered
as a necessary next step in the field of high-energy physics. If new
physics will show up at the CERN Large Hadron Collider (LHC), a LC
with a much cleaner environment would allow unambiguous precision
measurements. However the LC project has the potential to address,
on its own, questions of physics beyond the standard model, since
$e^{-}e^{-}$ and $\gamma \gamma$ options  are also planned beside
the basic $e^{+}e^{-}$ mode. If these options are carried on,
they will provide us for the first time with
the high physics potential of very high-energy $e^{-}e^{-}$ and
$\gamma \gamma$ collisions. See for example~\cite{ILC} for a full
discussion of the physics potential of the TESLA photon collider
(PC).

A topic which has recently received considerable attention is that
of neutrino mass and lepton number (flavor) violation, LNV (LFV).
Non-vanishing neutrino masses induce LFV processes such as $\ell
\rightarrow \ell' \gamma$. If neutrinos have masses in the eV or
sub-eV range, the neutrino generated branching ratio to the latter
process is of order ${\cal O}$(10$^{-40}$) and therefore
unobservably small. For such processes to be experimentally
accessible, new physics has to come into play.
Experimental searches of radiative lepton decays put strong bounds
on models of LFV:
$Br(\mu\to
e\gamma)<1.2\times 10^{-11}$~\cite{mue}, $Br(\tau\to
e\gamma)<3.9\times 10^{-7}$~\cite{taue}, $Br(\tau\to
\mu\gamma)<3.1\times 10^{-7}$~\cite{taumu}. Supersymmetric (SUSY)
extensions of the SM in the soft SUSY breaking potential $V_{soft}$
contains, in general, non diagonal entries in generation space and
therefore additional potential sources for LFV. Even in minimal
supergravity scenarios characterized by universal soft mass term for
scalar slepton and squark fields, renormalization induces
potentially sizable weak scale flavor mixing~\cite{Gabbiani} in
$V_{soft}$.

In this paper we study the lepton flavor violating reaction
\begin{equation}
{\gamma \gamma \to  \ell {\ell'}}
\label{gglfv}
\end{equation}
{with} $\ell \neq \ell'$ and $\ell,\ell' = e, \mu, \tau$, which
arises at one loop order in the just mentioned SUSY scenario, thus
extending to the $\gamma \gamma$ option an analysis done by some of
the authors in Ref.~\cite{cannoni} for the $e^{+}e^{-}$ and
$e^{-}e^{-}$ mode of the next linear collider. The OPAL
collaboration searched for this type of LFV reactions up to the
highest center-of-mass (CM) energy reached by LEPII, $\sqrt{s}= 209$
GeV~\cite{opal}. One $e^{+}e^{-}\to e\mu$ event was found at
$\sqrt{s}=189$ GeV matching all tagging conditions, but it was
interpreted as due to initial state radiation. These processes have
the advantage of providing a clean final state which is  easy to
identify experimentally (two back-to-back different flavor leptons),
though one has to pay the price of dealing with cross sections of
order $ {\cal O}(\alpha^4)$. In Ref.~\cite{cannoni} we found that
the $e^{-}e^{-}$ option stands better perspectives for the possible
detection of a LFV signal as opposed to the $e^{+}e^{-}$ mode, both
because of larger cross sections and smaller background. In general
the $\gamma \gamma$ mode offers larger cross section as compared to
the other modes, but at the same time
has the drawback of larger background and one must take into account
the non-monochromaticity of the beams.

The plan of the paper is the following: in
section~\ref{diagrams_and_amplitudes} we discuss the SUSY scenario
of LFV in the charged slepton sector (details of the helicity
amplitudes of the diagrams contributing to the signal reactions are
given in the appendix); in section~\ref{photon_spectra} we review
briefly the photon spectra used in the numerical computations of the
signal ; in section~\ref{results} we discuss the main features of
the signal; in section~\ref{background} we discuss the main standard
model (SM) backgrounds, and finally in section~\ref{conclusions} we
present the concluding remarks.

\section{SUSY scenario for lepton flavor violation }
\label{diagrams_and_amplitudes}

In the SUSY extension (with mSUGRA boundary conditions) of the seesaw
mechanism for the
explanation of neutrino masses~\cite{borma}, the superpotential
contains three $SU(2)_L$ singlet neutrino superfields $N_{i}$ with the following
couplings~\cite{borma,hisa2,hisa3}:
\begin{eqnarray}
W=(Y_{\nu})_{ij}\varepsilon_{\alpha\beta}H_{2}^{\alpha}N_i L^{\beta}_j
+\frac{1}{2}(M_{R})_i N_i N_i.
\label{yuk}
\end{eqnarray}
Here $H_2$ is a Higgs doublet superfield, $L_i$ are the $SU(2)_L$ doublet lepton superfields,
$Y_{\nu}$ is a Yukawa coupling matrix and $M_R$ is the $SU(2)_L$ singlet neutrino mass matrix.
At low energy
the renormalization group equations (RGE) produce within the MSSM diagonal
slepton mass matrices.
With the additional Yukawa couplings in Eq.~(\ref{yuk}) and the new mass
scale ($M_R$) the RGE evolution of the soft SUSY breaking parameters is
modified : assuming a {\em heavy} right handed singlet neutrino mass scale,
$M_{R}$, the RGE from the GUT scale down to $M_R$ induce
{\em off-diagonal} matrix elements in  $(m^{2}_{\tilde{L}})_{ij}$. In the one
loop approximation the off-diagonal elements are~\cite{hisa2}:
\begin{eqnarray}
(m^{2}_{\tilde{L}})_{ij}\simeq -\frac{1}{8\pi^{2}}(3+a^{2}_{0})m_{0}^{2}
(Y_{\nu}^{\dagger}Y_{\nu})_{ij}\ln\left(\frac{M_{GUT}}{M_{R}}\right).
\label{dl2}
\end{eqnarray}
$a_{0}$ is a dimensionless parameter appearing in the matrix of
trilinear mass terms
$A_{\ell}=Y_{\ell}a_{0}m_{0}$ contained in $V_{soft}$.
The rate of LFV transitions like $\ell_{i} \to \ell_{j}$, $i\neq j$,
$\ell=e,\mu,\tau$ induced by the lepton-slepton-gaugino vertex is determined by the
diagonalization  matrix ${U_{L}}_{ij}$.
These matrix elements can be potentially large because they are not directly
related to the mass of the light neutrinos,
but only through the seesaw relation $m_\nu \simeq m^2_D / M_R =v^2 Y^2_\nu/ M_R$.
The same effect on the mass matrix of $SU(2)_L$ singlet charged
sleptons $(m^{2}_{\tilde{R}})_{ij}$ is instead
much smaller : indeed,
in the same leading-log approximation of Eq.~(\ref{dl2}),
the corresponding RGE do not contain terms
proportional to $Y_{\nu}^{\dagger}Y_{\nu}$, since the right-handed
leptons fields only have
the Yukawa coupling $Y_{\ell}$, which completely determines the Dirac
mass of the charged leptons and these are known to be small numbers.
Thus the off-diagonal elements of $(m^{2}_{\tilde{R}})_{ij}$ can be taken
to a vanishing to a very good degree of accuracy.
The mixing matrix arising in the diagonalization of
$(m^{2}_{\tilde{L}})_{ij}$ induce LFV couplings in the
lepton-slepton-gaugino vertices
$\tilde{\ell}^{\dagger}_{L_{i}}{U_{L}}_{ij}\tilde{\ell}_{L_{j}}\chi$.
The magnitude of LFV effects will
in turn depend
on the fundamental theory in which this mechanism
is embedded (for example $SU(5)$ or $SO(10)$ SUSY
GUT~\cite{hisa3,biqi,masiso10}) and on the
particular choice of texture for the neutrinos
mass matrix~\cite{casas,ellis,pas}.

In this paper
we adopt a more general and phenomenological approach,
as it was done in~\cite{cannoni}, without
referring to a particular GUT model or neutrino mass texture.
We consider a two generation model for the  mass matrix of left-sleptons
(and sneutrinos):
\begin{eqnarray}
\widetilde{m}^{2}_{{L}}=\left(\begin{array}{cc}
\widetilde{m}^{2} & \Delta m^2\\
\Delta m^2 & \widetilde{m}^{2}
\end{array}\right),
\end{eqnarray}
with eigenvalues: $\widetilde{m}^{2}_{\pm}=\widetilde{m}^2\pm \Delta
m^2$ and maximal mixing. Under these assumptions, the  lepton flavor
violating propagator (in momentum space) for a scalar line is
\begin{eqnarray}
\langle\tilde{\ell}_{i}\tilde{\ell}^{\dagger}_{j}\rangle_{0}=
\frac{i}{2}\left(
\frac{1}{p^{2}-\widetilde{m}^{2}_{+}}-\frac{1}{p^{2}-\widetilde{m}^{2}_{-}}\right)
=i \frac{\Delta
m^2}{(p^{2}-\widetilde{m}^{2}_{+})(p^{2}-\widetilde{m}^{2}_{-})} \ ,
\label{LFVprop}
\end{eqnarray}
while the propagator for a lepton flavor conserving (LFC) scalar line is:
\begin{eqnarray}
\langle\tilde{\ell}_{i}\tilde{\ell}^{\dagger}_{i}\rangle_{0}=
\frac{i}{2}\left(
\frac{1}{p^{2}-\widetilde{m}^{2}_{+}}+\frac{1}{p^{2}-\widetilde{m}^{2}_{-}}\right).
\label{LFCprop}
\end{eqnarray}
The quantity
\begin{equation}
\delta_{LL}={\Delta m^{2}}/{\widetilde{m}^{2}} \label{delta}
\end{equation}
is the dimension-less parameter that controls the magnitude of the
LFV effect. This approach allows us to study  the signal in a quite
model-independent way by means of scans in the parameter space --
the ($\widetilde{m},\delta_{LL}$) plane -- which is already
constrained by the experimental bounds on radiative lepton decay
processes.

\begin{figure}[h]
\begin{center}
\scalebox{0.815}{\includegraphics*[10,320][550,840]{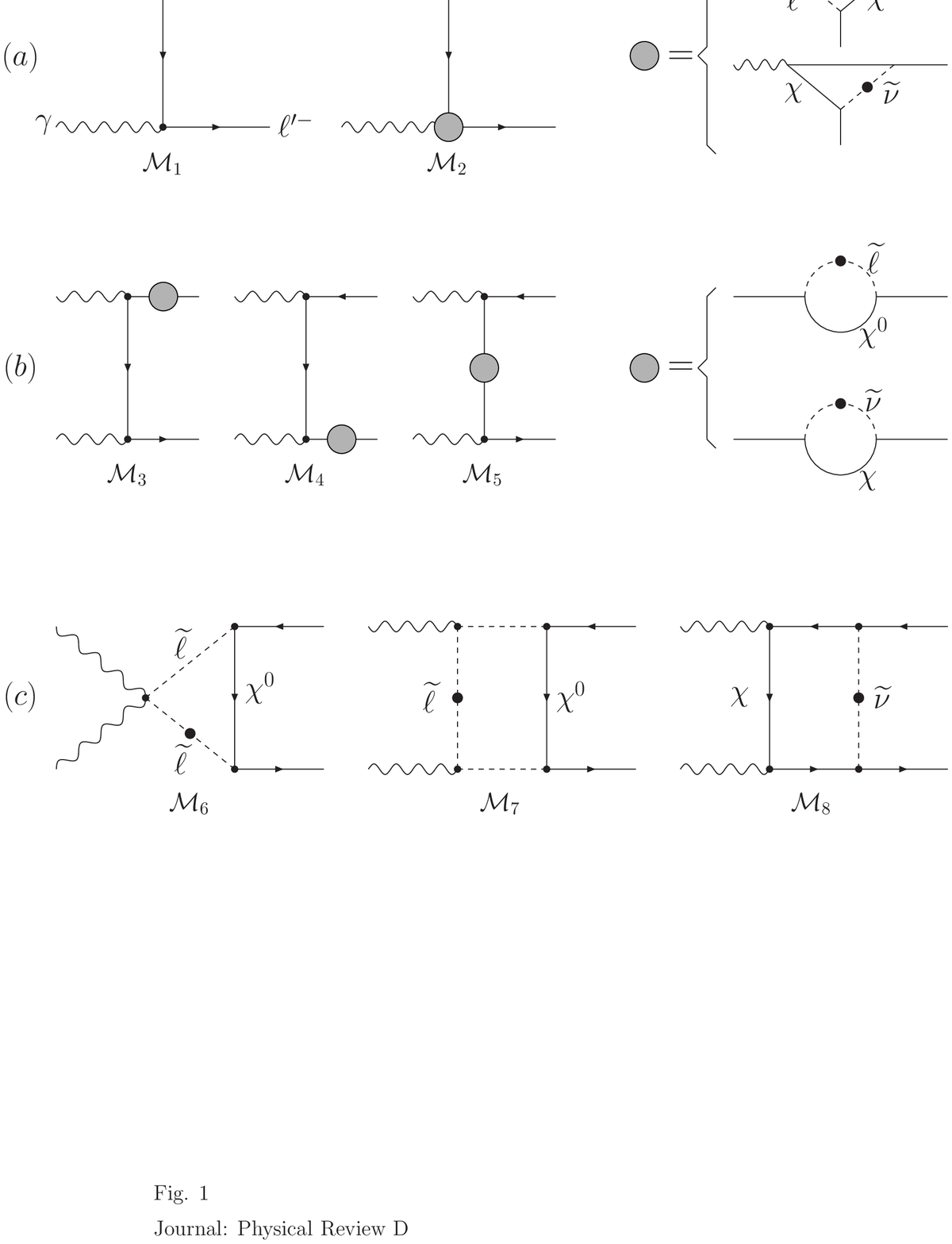}}
\caption{Diagrams for $\gamma \gamma$ collisions : $(a)$ penguin
diagrams ;  $(b)$ Self-energy diagrams ; $(c)$ box diagrams. The
full black dot in a scalar line denotes the insertion of the lepton
flavor violating propagator (Eq.~\protect\ref{LFVprop}). In the
diagrams of part $(c)$ this insertion is to be done successively in
each scalar line. In addition all graphs are accompanied by an
exchange graph where the final leptons are interchanged.}
\label{feynman_diagrams}
\end{center}
\end{figure}

For the calculations presented in this work it is a good approximation
to assume that the two lightest neutralinos are pure Bino and pure Wino
with masses $M_1$ and $M_2$ respectively, while
charginos are pure charged Winos with mass $M_2$, $M_1$ and $M_2$
being the gaugino masses in the soft breaking potential.
The Higgsino contribution to neutralino and charginos has suppressed amplitude,
since the coupling is proportional to the lepton masses.
For the same reason left-right mixing in the slepton matrix is neglected.
The relevant parts of the interaction lagrangian are,
adopting the notation of~\cite{Haber}:
\be
{\cal L}=
-g \ovl{\ell} P_R \widetilde{W} \tilde{\nu}+ \frac{g} {\sqrt 2}
\ovl{\ell} P_R \widetilde{W}^3 \tilde{\ell}_{L}+ \frac{g} {\sqrt 2}
t_W \ovl{\ell} P_R \tilde{B} \tilde{\ell}_{L}+ e \tilde{\ell}_{L}
A_{\mu} \tilde{L}^* i\olrap^{\mu} \tilde{\ell}_{L}- e A_{\mu}
\ovl{\widetilde{W}} \gamma^{\mu} \widetilde{W} \ee The contributing
one loop diagrams are displayed in Figure~\ref{feynman_diagrams}. We
have grouped them according to their topology : (a) penguin type ;
(b) self-energy ; (d) box diagrams. It is of course understood that
each diagram is accompanied by an exchange diagram where the final
state leptons (or initial state photons) are exchanged.

The possibility of having, at the next LC, high-energy polarized
photon beams suggests (see discussion in
section~\ref{photon_spectra}) to calculate the amplitudes of the
diagrams in Fig.~\ref{feynman_diagrams} within the helicity
formalism. Denoting by $\hat{T}$, $\hat{X}$, $\hat{Y}$, $\hat{Z}$
the space-time unit four-vectors, the four-momenta of the particles
in the center-of-mass frame (CMF) are expressed as:
\begin{equation}
\begin{array}{ll}
\displaystyle p_1 = \frac{\sqrt{s}}{2} \left(\hat{T} +
\hat{Z}\right)& \displaystyle \qquad p_3 = \frac{\sqrt{s}}{2}
\left(\hat{T} + \hat{X}\sin\theta^*
+ \hat{Z}\cos \theta^* \right)\\
\displaystyle p_2 = \frac{\sqrt{s}}{2} \left(\hat{T} -
\hat{Z}\right)& \displaystyle \qquad p_4 = \frac{\sqrt{s}}{2}
\left(\hat{T} - \hat{X}\sin\theta^* - \hat{Z}\cos \theta^* \right)
\end{array}
\label{momenta}
\end{equation}
where $s=(p_1 + p_2)^2$ and $\theta^*$ are, respectivley, the CMF
energy and scattering angle, while the polarization four-vectors of
the photons are:
\begin{equation}
\epsilon^\lambda_1 = -\frac{1}{\sqrt{2}}\left(\lambda \hat{X} +i \hat{Y}\right)
\qquad 
\epsilon^{{\lambda}^\prime}_2 = +\frac{1}{\sqrt{2}}\left({\lambda}^\prime \hat{X} - i\hat{Y}\right),
\label{helvect}
\end{equation}
where
$\lambda$ and ${\lambda}^\prime$ ($= \pm 1$) denote the photon helicities.
Assuming massless external fermions and,
given the chiral nature of the coupling
in the lagrangian, the helicity of the fermions in the final
state are fixed to
only one configuration, thus there are only four helicity
amplitudes corresponding to the
possible combinations of the photon helicities.
With obvious notation, we indicate with ${\cal M}^{(\lambda,\lambda')}$
the helicity amplitudes.
The loop integrals are decomposed in form factors according to the
notations of the software package
{\scshape{LoopTools}}~\cite{looptools} which is used in the code for
numerical computations. The final analytical formulas of the
amplitudes ${\cal M}^{(\lambda,\lambda')}$ are function of $s$,
$\theta^*$, $\lambda$, ${\lambda}^\prime$ and the SUSY parameters.
They also contain form factors originating from the loop integrals
which are defined according to the {\scshape{LoopTools}}
conventions. We report the explicit  expressions of the helicity
amplitudes in the Appendix~\ref{helicityamplitudes}.

\section{Discussion of photon beams and PC luminosity}
\label{photon_spectra}

High-energy photons beams~\cite{Ginzburg1,ILC} will be obtained
from Compton back-scattered (CB) low-energy laser photons
with energy $\omega_0$ off high-energy electron beams with energy $E_0$.
These high-energy photon beams
will not be
monochromatic but
will present instead an energy spectrum, mainly determined
by the Compton cross section, up to a maximum energy $y_m E_0$,
where $y_m =x/(x+1)$ with $x = 4 E_0 \omega_0/m^2_e$.

Full simulations of the experimental apparatus, see for example the simulation
of Telnov for TESLA~\cite{telnov}, show that the real luminosity spectrum
cannot be described by simple
analytical formulas
because of energy-angle correlation in Compton scattering, collisions effects and details of the
collision region.
Besides the high-energy peak also a 5-8 times higher low-energy peak is
present,
which is originated by photons after multiple Compton scattering and beamstrahlung
that cannot be described by analytical formulas.

The high-energy peak is instead
found to be almost independent of the technological details and well reproduced
by the product of two Compton spectra.
The normalized  Compton energy spectrum is:
\begin{equation}
F_c (x,y)\equiv \frac{1}{N_c} \frac{d N_c}{d y}=\frac{1}{N_c}\left[\frac{1}{1-y}-y+(2r-1)^2
-\lambda_e\, P_\ell\, x\, r\, (2r -1)(2-y)\right]
\label{spectrum}
\end{equation}
where $N_c$ is the normalization constant~\footnote{$
N_c =
\left[\left(1-\frac{4}{x}-\frac{8}{x^2}\right)\ln (x+1)+\frac{1}{2}+\frac{8}{x}-\frac{1}{2(x+1)^2}\right]
+\lambda_e P_l \left[\left(1+\frac{2}{x}\right)\ln(x+1)-\frac{5}{2}+\frac{1}{1+x}-\frac{1}{2(x+1)^2}\right]
$}, $y=E_\gamma /E_0$ is the fraction of the initial electron energy
acquired by the CB photons, $r=y/x(1-y)$,
and $\lambda_e$ and $P_l$
are the electrons and laser photons polarizations
($|\lambda_e, P_l|\le 1$), respectively.
Thus the theoretical differential spectrum for luminosity is:
\be
\frac{d L_{\gamma\gamma}^{CB}}{d y_1 d y_2}=
\,F_c (x,y_1 )\,F_c (x,y_2).
\label{ideallum}
\ee
It is useful to rewrite Eq.~(\ref{ideallum}) in terms of the
invariant variables
$z=\sqrt{y_1 y_2} = W_{\gamma \gamma} /2E_0
= \sqrt{s_{\gamma \gamma}/s_{ee}}$
and the
pseudo-rapidity
$\eta = \ln{\sqrt{y_1/y_2}}$, and define  a
differential spectrum as a function of $z$ :
\be
\frac{d L_{\gamma\gamma}^{CB}}{d z}=
2 z \,\int^{\ln{y_m /z}}_{-\ln{y_m /z}} F_c (x,z e^{+\eta}) F_c (x,z e^{-\eta})d\eta .
\label{ideallum1}
\ee
\begin{figure}
\begin{center}
\scalebox{0.5}{\includegraphics*[40,130][540,710]{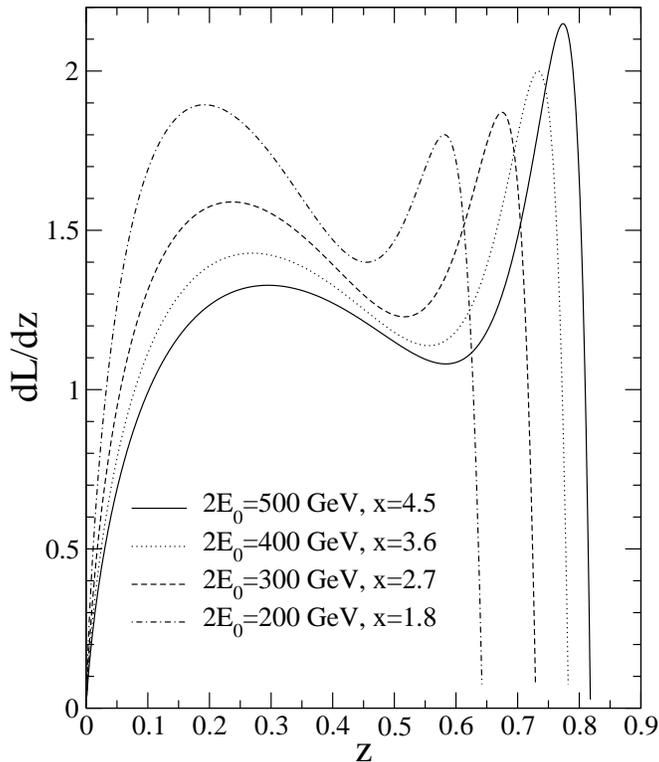}}
\caption{The ideal CB luminosity spectrum plotted for different CM energies
of the $ee$ collider.
}
\label{lumspectra}
\end{center}
\end{figure}
This is the  function we have plotted in  Fig.~\ref{lumspectra}
for some values of $E_0$ and for correlated value of $x$ calculated
using the TESLA parameter $\omega_0 = 1.17 $ eV.
It gives a peak of luminosity near the maximum value of $z$,
$z_m = W_{\gamma\gamma\, \text{max}} /2E_0 = y_m$,
as shown in Fig.~(\ref{lumspectra}) and a broad spectrum at lower values.
This means that most of the collisions involve two high-energy photons from
the high-energy peaks of the two Compton spectra.
We note that the peak in the luminosity spectrum is obtained when the product $P_\ell\,\lambda_e$
is negative for both Compton spectra.

In this high-energy range, the colliding photons have practically the
same energy which is close to its maximum
value. Obviously, this configuration is the most favourable to
distinguish two-particle final states
among multi-particle production. It is important to notice that
the experimental design for a future
photon-photon collider is planned so as to have full control of
the luminosity and optimize it
in this high-energy range in view of Higgs physics studies~\cite{telnov}.
In the low-energy range, collisions between photons that may have
very different energies take place, leading to copious boosted
events. Then, the separation between signal and background becomes
more challenging. Moreover, this low-energy part is more dependent
of the experimental apparatus. For these reasons, we have restricted
our study to the  high-energy part  of the luminosity spectrum.
Another reason to restrict the peak is that the total luminosity of
the photon collider is defined by the condition \be
L_{\gamma\gamma}=\int^{z_{max}}_{0.8z_{max}}dz\frac{dL_{\gamma\gamma}}{dz}.
\ee To evaluate the expected total number of events and event rates
we take as benchmark the TESLA parameters in Ref.~\cite{ILC}. At
$\sqrt{s_{ee}}=2 E_0= 200$, $500$ GeV the geometrical luminosities
are expected to be $L_0 =$ ($4.8$, $12$, $19.1$)$\times 10^{34}$
cm$^{-2}$ s$^{-1}$, while the corresponding photon-photon
luminosities at the peak are : $L_{\gamma\gamma}(z>0.8z_m)=$(0.44,
1.15) $\times 10^{34}$ cm$^{-2}$ s$^{-1}$, equivalent to (1.3, 3.4)
$\times$ 10$^{2}$ fb$^{-1}$ yr$^{-1}$. To use these simulated
realistic numbers with the ideal spectrum in Eq.~(\ref{ideallum1})
a suitable normalization is necessary : \be
L_{\gamma\gamma}=C_{norm}\int^{z_{max}}_{0.8z_{max}}~2z~dz
\,\int^{\ln{y_m /z}}_{-\ln{y_m /z}} F_c (x,z e^{+\eta}) F_c (x,z
e^{-\eta})d\eta . \label{lnorm1} \ee \be
N_{events}=L_{\gamma\gamma}\int^{z_{max}}_{0.8z_{max}}dz\frac{1}{L_{\gamma\gamma}}
\frac{dL_{\gamma\gamma}}{dz}\sigma(W_{\gamma\gamma}) \label{lnorm2}
\ee Substituting Eqs.~(\ref{lnorm1},\ref{lnorm2}) into the integral
we eliminate the dependence  from $C_{norm}$, which depends on the
total integrated luminosity, redefining the differential spectrum as
\be
\frac{dL^{norm}_{\gamma\gamma}}{dz}=L_{norm}\frac{dL^{CB}_{\gamma\gamma}}{dz},
\ee where $L_{norm}$ is given by: \be
L_{norm}&=&\frac{1}{\int^{z_{max}}_{0.8z_{max}}dz 2z \,\int^{\ln{y_m
/z}}_{-\ln{y_m /z}} F_c (x,z e^{+\eta}) F_c (x,z e^{-\eta})d\eta}
\ee Thus we define both for signal and background the {\it
effective} cross section as : \be
\sigma^{effective}=\int^{z_{max}}_{z_{min}}dz
\frac{dL^{norm}_{\gamma\gamma}}{dz} \sigma(W_{\gamma\gamma}) \ee and
the total number of events is thus given by
$N_{events}=L_{\gamma\gamma}\times\sigma^{effective}$.
\begin{figure}
\begin{center}
\scalebox{0.5}{\includegraphics*[40,115][570,710]{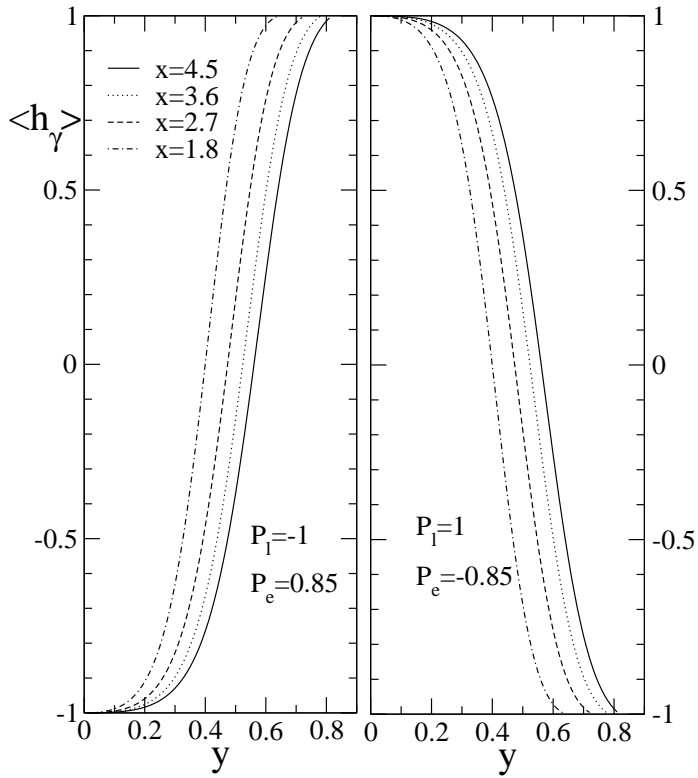}}
\caption{The ideal CB helicity spectrum plotted for different center
of mass energies of the $ee$ collider.
}
\label{elicita}
\end{center}
\end{figure}
In view of
studying helicity correlations,
we discuss the polarizations properties
of the back-scattered photons.
The degree of circular polarization is given by:
\be
\langle h_{\gamma}\rangle=\frac{-P_{\ell}(2r-1)\left[(1-y)^{-1}+1-y\right]
+2\lambda_e xr\left[1+(1-y)(2r-1)^{2}\right]}{N_c \;F_c (x,y)}
\label{elicitamedia}
\ee
Assuming complete polarization for laser photons ($P_\ell = \pm 1$) and the planned
maximum available for electrons $\lambda_e =\pm 0.85$, this function is plotted in
Figure~\ref{elicita} for various values of $x$.
As can be seen,
in the high-energy peak where $y$ is near $y_m$, colliding photons
have a high degree of circular polarization 
with $P_{\gamma}=-P_l$.

\section{Discussion of the signal}
\label{results}
We discuss first the signal for the ideal case with monochromatic
photons in pure helicity state.
The differential polarized cross sections with respect to the scattering angle
in the photon-photon CMF are given by
\be \frac{d{\hat{\sigma}}^{\lambda\lambda'}}{d\cos{\theta}^*}=
\frac{1}{32\pi \hat{s}} \big| {\cal M}^{\lambda\lambda'}
(\hat{s},\hat{t},\hat{u})\big|^2\,. \label{diff1} \ee
\begin{figure}
\begin{center}
\scalebox{0.5}{\includegraphics*[4,115][510,710]{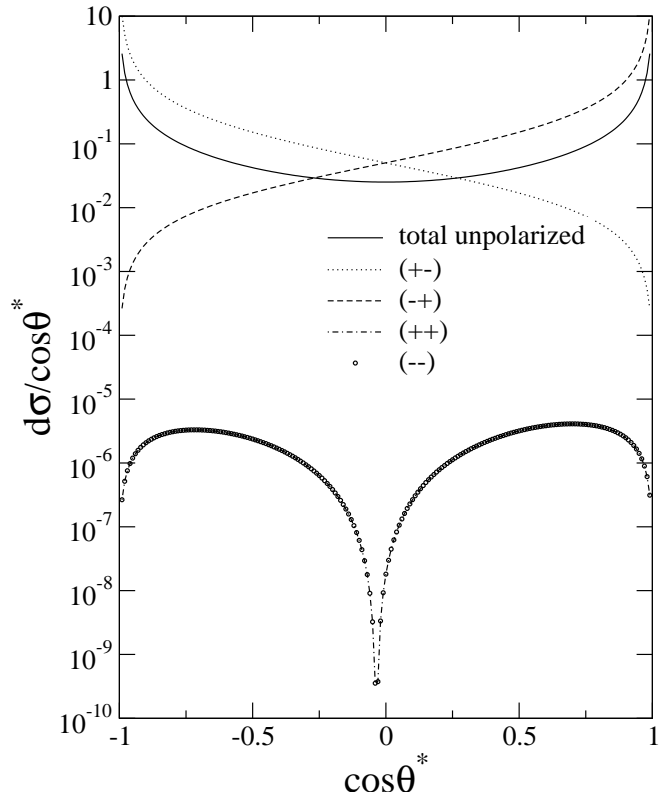}}
\caption{Differential cross section given by the four helicity
amplitudes for monochromatic photons at $\sqrt{s_{\gamma\gamma}}
=128$ GeV. The values of the masses are $M_1 =200$, $M_2 =100$,
$\langle \tilde{m}_l \rangle = 150$ GeV and $\Delta m^2 = 6000$
GeV$^2$.} \label{angolar}
\end{center}
\end{figure}
We plot them in Figure~\ref{angolar} as functions of the CMF
scattering angle with masses set to the values specified in the
caption of the figure and for $\sqrt{s_{\gamma\gamma}} =128$ GeV
which corresponds to the maximum energy that is reachable with a LC
with $\sqrt{s_{ee}} =200$ GeV.
It is seen that the amplitudes with opposite helicity photons
${\cal M}^{(+,-)}$ and ${\cal M}^{(-,+)}$ ($J_z = \pm 2$)
dominate the signal,
while those with same helicity photons ($J_z = 0$) give negligible cross-sections.
Moreover the former are peaked in the forward and backward directions while the second
are suppressed in these regions.
\begin{figure}
\begin{center}
\scalebox{0.5}{\includegraphics*[20,100][530,710]{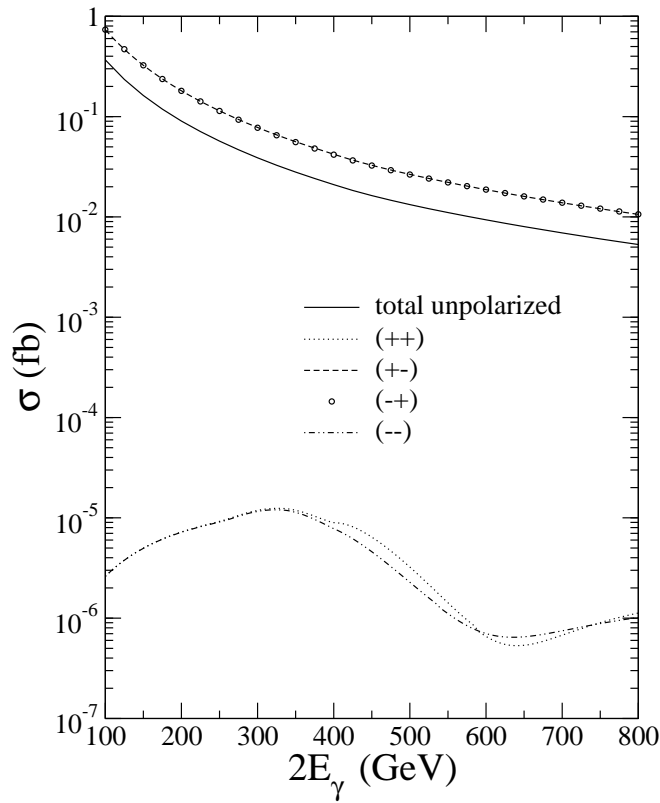}}
\caption{Total signal cross sections for monochromatic photons as a
function of the energy with the parameters as specified in the
caption of Figure~\ref{angolar} .} \label{totale}
\end{center}
\end{figure}
The total cross sections are plotted in Figure~\ref{totale} varying the CM energy.
The $J_z = \pm 2$ cross sections decrease with increasing energy for they are dominated by
the  diagrams with the exchange of a {\em light} lepton in the
$t$ and $u$ channels [(a) and (b) of Fig.~\ref{feynman_diagrams}].

The realistic effective differential cross sections as function of
the scattering angle in the laboratory ($e^-e^-$ CMF) are simply obtained by a boost
and by
convoluting the ``monochromatic'' differential
cross-sections in Eq.~(\ref{diff1}) with the luminosity spectrum
discussed in the preceding section. The fact that photons are not
in pure helicity state is here taken into account using density matrices for initial photons
expressed in terms of Stokes parameters~\cite{landau}.
The complete formula is:
\begin{equation}
\frac{d{{\sigma}}^{\lambda\lambda'}}{d\cos{\theta}}=
\int^{z_{max}}_{z_{min}}dz\,\, 
\frac{dL^{norm}_{\gamma\gamma}}{dz}
\frac{d\cos{\theta}^*}{d\cos{\theta}}
\frac{\left(1-\langle\lambda(x,P_{\ell_1},\lambda_{e_1})\rangle\right)}{2}
\frac{\left(1-\langle\lambda'(x,P_{\ell_2},\lambda_{e_2})\rangle\right)}{2}
\frac{d{\hat{\sigma}}^{\lambda\lambda'}}{d\cos{\theta}^*}
\label{diff2}
\end{equation}
Here the functions $\langle\lambda(x,P_{\ell_1},\lambda_{e_1})\rangle$,
$\langle\lambda'(x,P_{\ell_2},\lambda_{e_2})\rangle$
play the role of the Stoke parameter
$\eta_2$, while $\eta_1$ and $\eta_3$~\cite{landau}
give no contribution for we are assuming laser photons with full circular
polarization.
The total cross sections are obtained finally by integrating over the
laboratory scattering angle and introducing a kinematical cut :
\begin{equation}
\sigma^{\lambda\lambda'}(s_{ee})=\int^{(\cos\theta)_{max}}_{(\cos\theta)_{min}}d\cos\theta\,
\frac{d{\hat{\sigma}}^{\lambda\lambda'}}{d\cos{\theta}}
\end{equation}

\begin{figure}
\begin{center}
\scalebox{0.5}{\includegraphics*[20,20][520,760]{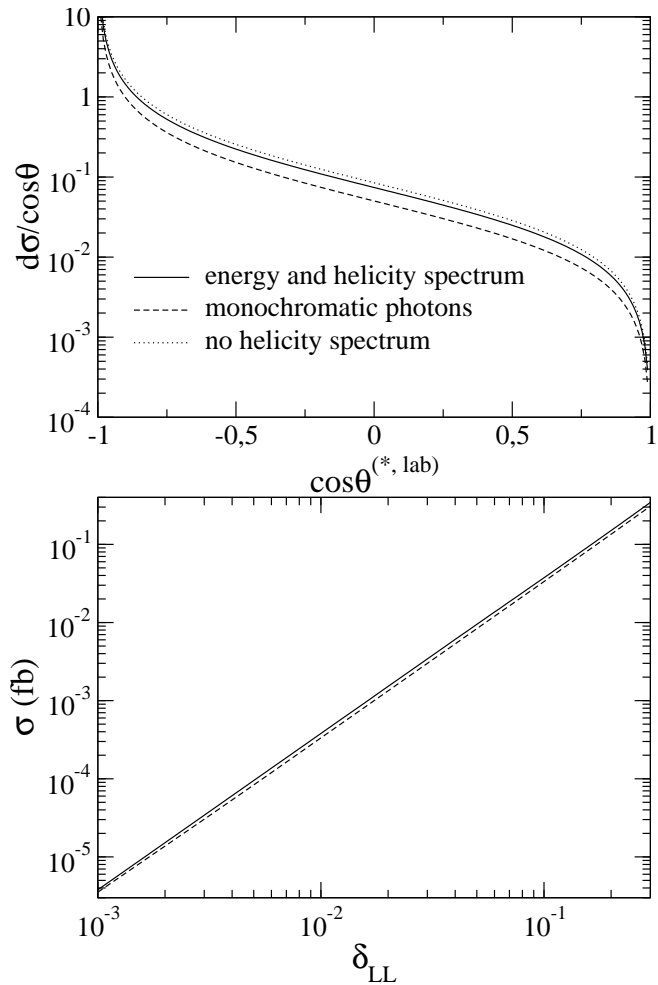}}
\caption{Effect of energy and helicity spectra on angular and total
cross sections for the helicity configuration (+ -). The values of
masses are the same as in Figures~\ref{angolar}, ~\ref{totale}. }
\label{effectoncross}
\end{center}
\end{figure}
In Figure~\ref{effectoncross} we study the effect of the inclusion
of spectra in the calculation. The upper panel presents the same
monochromatic (+ -) differential cross section of
Figure~\ref{angolar} compared with the one calculated with the
complete formula of Eq.~\ref{diff2}, while the bottom panel shows
the total cross section as a function of the parameter $\delta_{LL}$
(c.f. Eq.~\ref{delta}) :
it is clear that the complete formula in Eq.~\ref{diff2} gives
results almost identical (within a few $\%$) to the monochromatic
calculation with photon energies fixed to their maximum values. This
is a consequence of the choice of restricting the calculation to the
luminosity peak near $z_m$ where the photons have energies near
$E_{\gamma}^{max}$, are in an almost defined helicity state and the
boost to the lab
system has $\beta\simeq 0$. Thus, in the following we consider the
cases of a photon collider with $2E_0 = 200$ and 500 GeV, with
monochromatic photons in pure helicity states with
$E_{\gamma}=E_{\gamma}^{max}$ ($\sqrt{s_{\gamma\gamma}} = 128$ and
410 GeV
respectively)
 and use the realistic simulated
luminosities of TESLA to estimate event rates.

\begin{figure}
\begin{center}
\scalebox{0.5}{\includegraphics*[32,90][500,800]{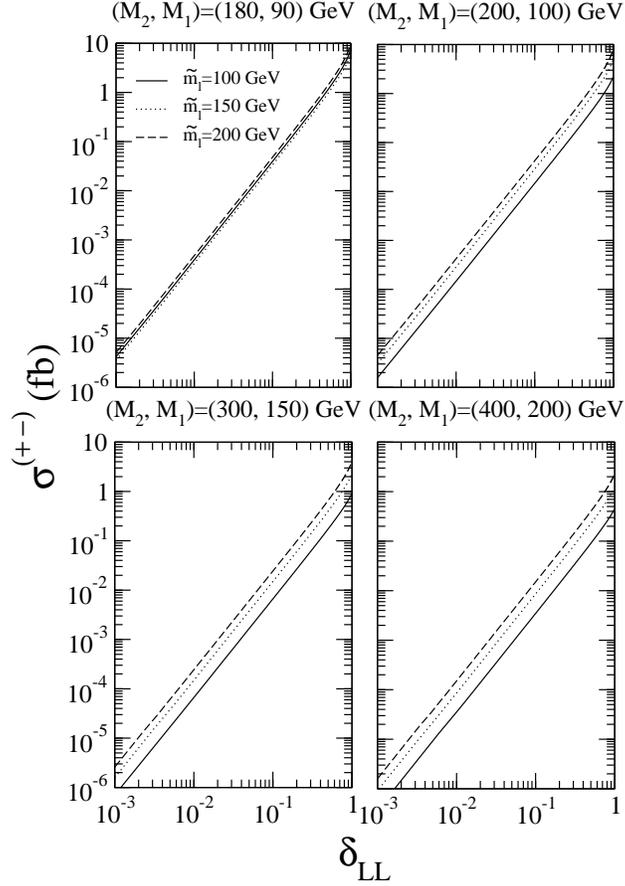}}
\caption{Total cross section for the amplitude (+ -) as a function
of the dimensionless parameter $\delta_{LL}$ and
$\sqrt{s_{\gamma\gamma}} =128$ GeV. The values of the others
parameters are given in the legends. } \label{sd200}
\end{center}
\end{figure}

\begin{figure}
\begin{center}
\scalebox{0.5}{\includegraphics*[32,90][500,800]{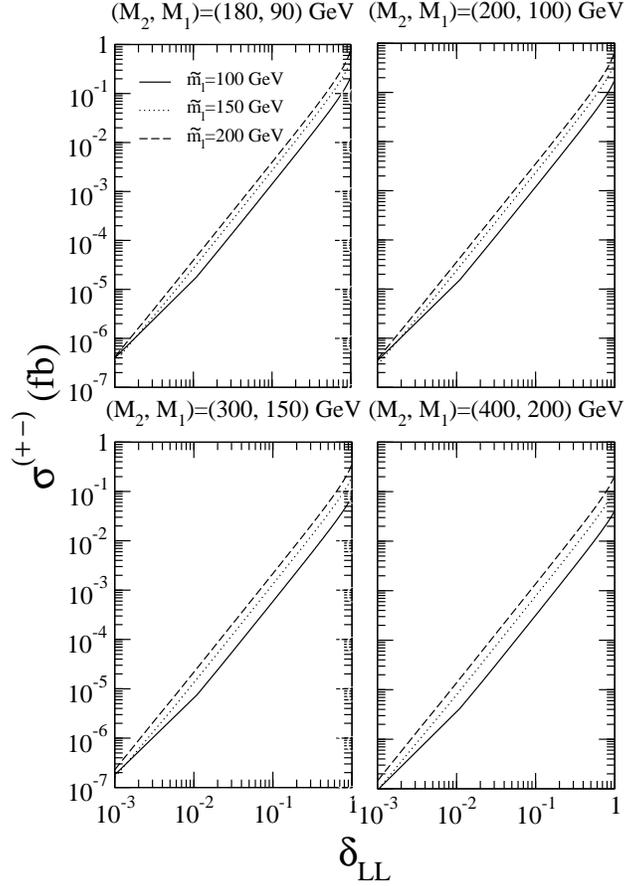}}
\caption{Total cross section for the amplitude (+ -) as a function
of the dimensionless parameter $\delta_{LL}$ and
$\sqrt{s_{\gamma\gamma}} =410$ GeV. The values of the others
parameters are given in the legends. } \label{sd500}
\end{center}
\end{figure}

In Figures~\ref{sd200} and~\ref{sd500} we plot the cross section
given by the dominant amplitude (+ -) as function of the insertion
$\delta_{LL}$ for some values of gaugino and average slepton masses.
These values are in the range of the SPS1 benchmark point mSUGRA
scenario~\cite{sps} that give a particle spectrum with the lightest
charginos, neutralinos and sleptons in the $100-200$ GeV region.
This light spectrum is also favoured by global fits to the standard
model parameters~\cite{altarelli}. Even if the differential cross
section is peaked along the collision axis, a necessary  angular cut
$|\cos \theta | <0.9$ is applied because the background is also
large in this region, as it is discussed in
Section~\ref{background}. Given luminosities of order ${\cal O}
(100)$ fb$^{-1}$yr$^{-1}$, cross sections greater than $10^{-2}$ fb
are needed. In the case of a $2E_0 = 200$ GeV~ PC, this happens for
$\delta_{LL} > 5\times 10^{-2}$ while in the $2E_0 = 500$ GeV case,
$\delta_{LL} > 5\times 10^{-1}$ is needed, which means quite large
non-diagonal matrix elements.

To see if these large mass splittings are allowed by current
experimental constraints we have to take into account the bounds
imposed on the model by the non observation of radiative decays.
\begin{figure}
\begin{center}
\scalebox{0.5}{\includegraphics*[50,15][550,815]{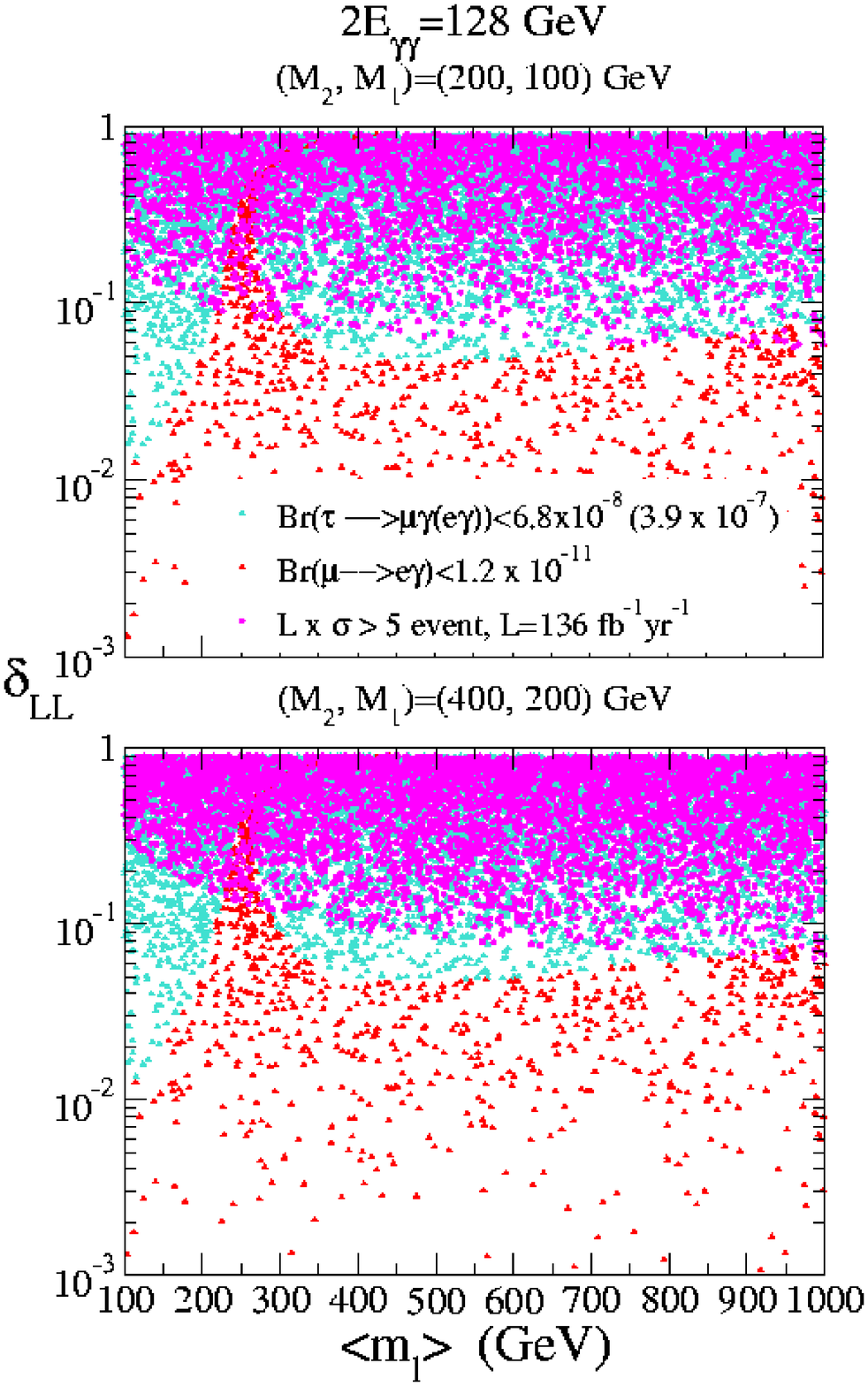}}
\caption{Scatter plot in the plane ($\delta_{LL}, \widetilde{m}$)
of: (a) the experimental bounds from $\mu\to e\gamma$ and $\tau \to
\mu \gamma$ (allowed regions with circular dots); (b) regions where
the signal can give at least five events at year for two sets of
gaugino masses. The energy is $\sqrt{s_{\gamma\gamma}} =128$ GeV and
the luminosity $L=136$ fb$^{-1}$ yr$^{-1}$. } \label{sca200}
\end{center}
\end{figure}
\begin{figure}
\begin{center}
\scalebox{0.5}{\includegraphics*[50,15][550,815]{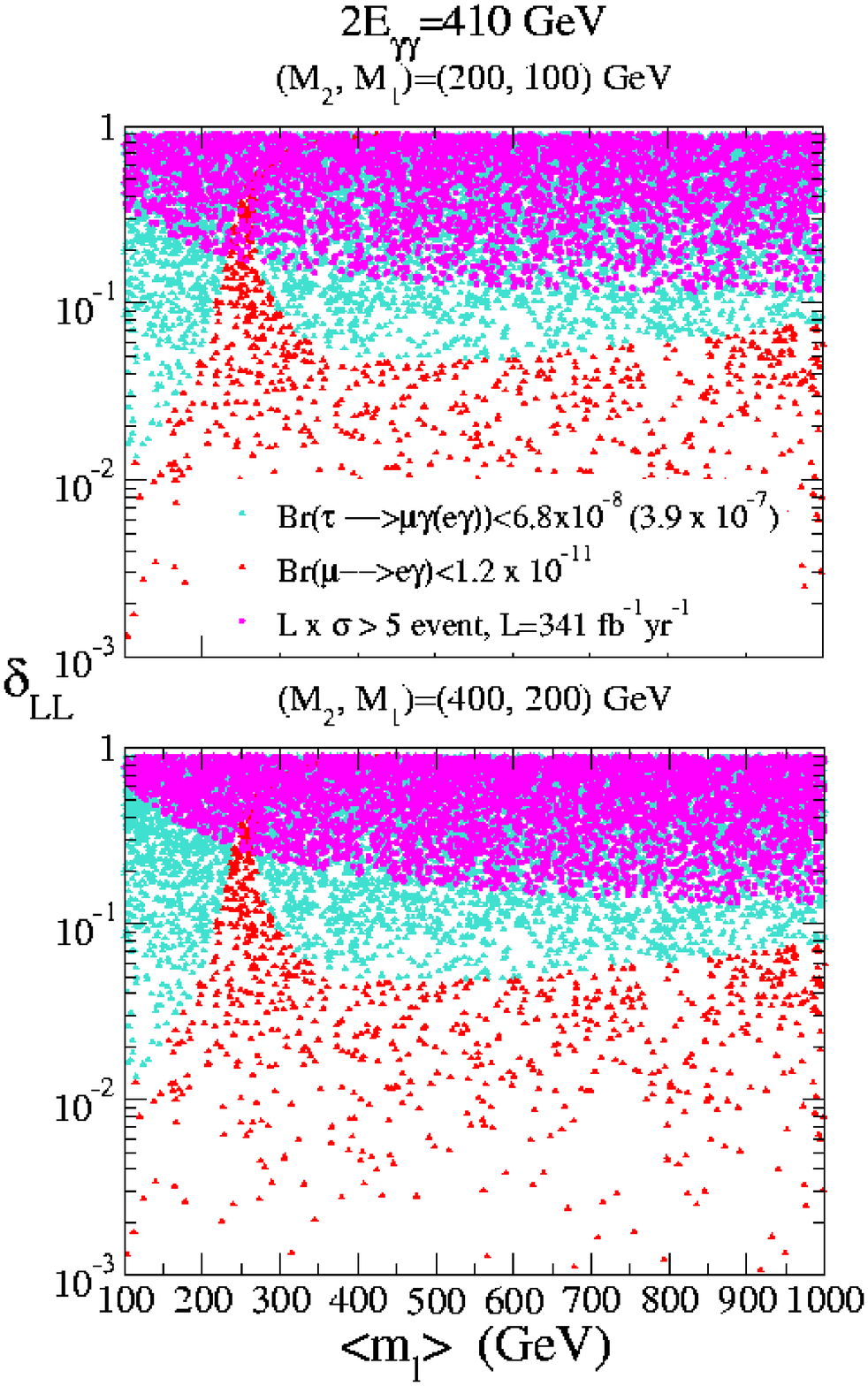}}
\caption{Scatter plot in the plane ($\delta_{LL}, \widetilde{m}$)
of: (a) the experimental bounds from $\mu\to e\gamma$ and $\tau \to
\mu \gamma$ (allowed regions with circular dots); (b) regions where
the signal can give at least five events at year for two sets of
gaugino masses. The energy is $\sqrt{s_{\gamma\gamma}} =410$ GeV and
the luminosity $L=341$ fb$^{-1}$ yr$^{-1}$. } \label{sca500}
\end{center}
\end{figure}
In Figures~\ref{sca200},~\ref{sca500} we show scatter plots where
the average lepton masses and the relative mass splitting
$\delta_{LL} = \Delta m^2 /\widetilde{m}^2$ are varied freely, for
fixed values of gaugino masses. All the parameter space [the
$(\widetilde{m}, \delta_{LL})$ plane] is covered by the clear
triangle-up shaped points (turquoise in color) that satisfy the
bounds $\Gamma (\tau \to \gamma \mu,( e))<6.8\times 10^{-8}\;3.9
\times 3.9\times 10^{-7} $
while the 
black triangle-up shaped points (red in color), that satisfy $\Gamma
(\mu \to e\gamma) <1.2\times 10^{-11}$ cover a more restricted part.
The grey circle shaped  points (magenta in color) are determined
imposing the condition that the total cross section multiplied by
the luminosity gives more than five events per year. We can note two
things: the signal's points overlap with the ``$\mu,e$'' region only
on the tail of the red region extending to higher values of
$\delta_{LL}$. This tail is due to some peculiar cancellation
between diagrams, as discussed in Ref.~\cite{hisa2}, thus we can say
that this possible final state is almost excluded. The $\mu, \tau$
and the $e, \tau$ final state are not excluded but they generally
require a high-mass splitting $\delta_{LL} > 10^{-1}$.

From the point of view of the supersymmetric seesaw mechanism
described in Section~\ref{diagrams_and_amplitudes}, these values can
be realized in nature only under some restricted
conditions~\cite{pascoli} : the matrix $Y_{\nu}$ from the seesaw
mechanism neutrino masses and mixing is ambiguous up to a complex,
orthogonal matrix $R$~\cite{casas}. Usually this matrix is taken to
be real or identical to the unit matrix. In the case of a
quasi-degenerate neutrino mass spectrum, $R$ being complex allows
for values of $\Delta m^2$ larger by 5-8 orders of magnitude
relative to the case of $R$ being real or the unit
matrix~\cite{cannoni}.

\section{Standard Model background}
\label{background}

Production of charged leptons will be copious in $\gamma\gamma$ collisions,
and the  SM provides  several processes that can mimic  $e\tau$,
$\mu\tau$ final states.
Let us see how to reduce the most important contributions that are :
\begin{eqnarray}
&&(a)\;\;\gamma\gamma \to  \tau^{-} \tau^{+} \to \tau^-
\nu_e \bar{\nu}_{\tau} e^+
\label{tt}\\
&&(b)\;\;\gamma\gamma \to {W^{-}}^{*} {{W}^{+}}^{*} \to \tau^{-}
{\bar{\nu}}_{\tau}\,{e^{+}} {{\nu}}_{e}
\label{ww}\\
&&(c)\;\;\gamma\gamma \to  e^+ e^- \tau^+ \tau^-
\label{et}
\end{eqnarray}
with similar processes for the production of $\mu\tau$ pairs. As we
have seen, the $e\mu$ final state, which is the easiest to
reconstruct from the experimental point of view, is almost
completely excluded by the strong bounds from the non observation of
the radiative decay $\mu \to e \gamma $.  Thus we are bound to
consider signals with a tau in the final state. Taus can, in
principle, be reconstructed looking at the associated leptonic decay
$\tau \to \ell \nu \bar{\nu}$ and at the hadronic decay $\tau \to
\pi^\pm \pi^0$. The cross sections of processes in
Eqs.~(\ref{tt}-\ref{ww}) depend on initial photon polarizations,
while the reaction in Eq.~(\ref{et}) is almost insensitive to photon
helicities. We use the program {\scshape{COMPHEP}}~\cite{comphep},
and the CB spectra with $z>0.8z_{max}$. In Table~\ref{table1} we
give  the values of the cross sections after the application of
kinematical cuts (contributions of the charge-conjugate processes
are also included).
\begin{table}[t]
\begin{ruledtabular}
\caption{
Total cross section without and with cuts described in the text for the background processes
Eqs.~(\ref{tt}-\ref{et}).
}
\begin{tabular}{cc|c|c|c}

2$E_0$ (GeV) &   & $\gamma\gamma \to \tau\tau \to\tau e\nu\bar{\nu}$ \ \ \ \ \ \ \ \ & $\gamma\gamma \to WW \to e\tau\nu\bar{\nu}$
\ \ \ \ \ \ \ \ & $\gamma\gamma \to \tau\tau ee$ \ \ \ \ \ \ \ \ \cr
\hline\hline

200 & $\sigma$       & \ \ \ 0.58 fb  & 2.3$\times10^{-1}$   &
36.7 pb
                      \cr
    & $\sigma_{cut}$ \ \ \ & 1.49$\times 10^{-6}$ fb  &  //
 & 4.4$\times 10^{-2}$ fb
     \cr

\hline

300 & $\sigma$       & 3.1 fb  & 0.48 pb    &
38.9 pb
\cr
    & $\sigma_{cut}$ & 16.3$\times 10^{-6}$ fb  &  //  &
3.7$\times 10^{-2}$ fb
        \cr
\hline

400 & $\sigma$       & 4.9 fb    & 0.69 pb  &
39.5 pb
 \cr
    & $\sigma_{cut}$ & 3.9$\times 10^{-4}$ fb     &  $2.1\times 10^{-2}$ fb &
2.9$\times 10^{-2}$ fb
      \cr

\hline
500 & $\sigma$       & 6.1 fb  & 0.77 pb &
39.9 pb
\cr
    & $\sigma_{cut}$ & 9.7$\times 10^{-4}$ fb    &  $1\times 10^{-1}$ fb  &
2.4$\times 10^{-2}$ fb

\label{table1}
\end{tabular}
\end{ruledtabular}
\end{table}
Tau pair production,
$W$ gauge bosons pair production and four charged fermion production
are known to have very large cross sections, at the level of hundreds
of picobarn at the
CM energy of ILC, orders of magnitude larger than the
signal in the most favourable regions of the parameter space.
However the signal is characterized by  two back-to-back leptons
with the energy of the beams {\em without missing transverse momentum
and energy}. These
characteristics provide also  indications on the necessary kinematical
cuts to be applied to the background processes.

The helicity amplitudes which dominate the signal $(+,-)$ and $(-,+)$,
are peaked along the collision axis.
Most of the background is also concentrated in this region. So we
apply the angular cut $|\cos(\theta)|<0.9$ ($\theta < 25.8^{\circ}$)
both to the signal and to the background. We also impose the
back-to-back condition on the background processes, requiring
$180^{\circ} - \theta_{\ell \ell '} <5^{\circ}$. Using in addition
the condition that one of the event hemisphere should consist of a
single muon or electron with energy close to $E_\gamma$, final
leptons are required to have energy at least $85\%$ of the maximum
photon energy $E^{\gamma}_{max} =y_{max}E_0$.

As can be seen from Table~\ref{table1}, after these cuts are
applied, process $(a)$ is suppressed because tau pairs are almost
produced along the collision axis, and process $(b)$ is completely
excluded, at least for low energies, because the leptons from the
decay of $W$ are less energetic and cannot survive to the energy
cut.
Instead, due to the well known rapid growth of the $\gamma \gamma
\to WW$ cross section above threshold, at $400$ and $500$ GeV CMF
energy, these cuts are not enough to suppress the background, giving
cross sections of $2.1\times 10^{-2}$ fb and $10^{-1}$ fb
respectively. However with a cut on the transverse momentum of the
electron $p_T^e
>15$ GeV the cross section, at $\sqrt{s_{ee}} = 500 $ GeV, is reduced to
$2.1\times 10^{-2}$ fb, while for $p_T^e >20 $ GeV, the contribution
is eliminated.

Reaction $(c)$ turns out to provide the most dangerous background.
In this case the results were obtained
with a MonteCarlo code developed by some of the
authors~\cite{carimalo}, which  uses some compact analytical
expressions for the diagrams with the exchange of space-like
photons. The configuration that mimics the signal arises if one
$e\tau$ pair is emitted at small angle with respect the collision
axis and {\em is not} detected (we require $\theta_\ell^{untagged} <
25.8^{\circ}$), while the other pair is tagged. This configuration
is determined by multi-peripheral diagrams as a consequence of
$t$-channel poles at small angles. The detected pair presents
characteristics very similar to those of the signal, and though the
cross section is effectively reduced by orders of magnitudes, it is
still at the level of $10^{-2}$ fb, thus remaining competitive with
the signal cross section.
However, at a final step, this background can be estimated from the
data by requiring instead that the detected tau and electron be of
the {\em same} charge, and eventually subtracted.
After the cuts discussed above this is the only significant
background contribution which remains. We consider the statistical
significance
\[SS=\frac{{\cal L} \sigma_{cut}^{S}}{\sqrt{{\cal L} \sigma_{cut}^{BG}}}\]
and requiring $SS\ge 3$ we obtain $\sigma_{cut}^{S}>5.4\times
10^{-2}$ fb at $\sqrt{s_{ee}} = 200$ GeV and
$\sigma_{cut}^{S}>2.5\times 10^{-2}$ fb at $\sqrt{s_{ee}} = 500$
GeV, using the simulated annual luminosity for TESLA. By inspection
of Fig.~\ref{sd200} and Fig.~\ref{sd500} it is seen that this
condition   is in both cases  satisfied  if $\delta_{LL}\gtrsim
10^{-1}$ with the values of the other SUSY parameters as specified
previously. This region of the parameter space is allowed for the
$e\tau$, $\mu\tau$ channels as can be seen in Fig.~\ref{sca200} and
Fig.~\ref{sca500}.

\section{Summary and conclusions}
\label{conclusions}

We have studied the lepton flavor violating reactions ${\gamma
\gamma \to  \ell \ell'}$ (${\ell,\ell^\prime = e, \mu,\tau}$,
$\ell\ne\ell'$) which arise at the one loop order of perturbation
theory and which will be of interest for the $\gamma\gamma$ option
of the future ILC. The LFV mechanism is provided by low energy
R-conserving supersymmetry with non diagonal slepton mass matrices.
The origin of the non diagonal entries of the charged slepton mass
matrices can be ascribed to a SUSY seesaw mechanism with mSugra
boundary conditions, a theoretical scenario that has attracted much
attention in the literature in recent years. We have studied the
signal in a model independent way in order to pin down regions of
the SUSY parameter space, the ($\widetilde{m}_\ell , \delta_{LL}$)
plane, allowed by the present experimental limits.

We have shown that in the range 200-500 GeV for the center of mass energy
of the basic electron collider that produces photon beams,
the cross section of the signal is $\sigma(\gamma\gamma\to \ell \ell') =
{\cal O}(10^{-1} \, -\,  10^{-2})$ fb for sparticle masses in the range
$90-200$ GeV that correspond to a light SUSY spectrum somehow hinted to by
fits on standard model parameters and SUSY benchmark points.
Observation at a PC of  $\gamma \gamma \to e\tau$, $(\mu\tau)$
is not excluded by present bounds on the radiative lepton decays
$\tau \to e\gamma$, $\tau \to \mu\gamma$ which do not constrain the
parameter space strongly enough, nonetheless
a relative mass splitting
$\delta_{LL}={\Delta m^{2}}/{\widetilde{m}_\ell^{2}}$
at least of order $10^{-1}$ is required,
a value that can be obtained in the SUSY
seesaw framework but only within some particular model.
The $e\mu$ final state is almost excluded because of the stronger
constraint provided  by the upper bound on the branching ratio
given by the non observation of $\mu \to e\gamma$ which is four orders of
magnitude smaller than those provided by the non observation of
$\tau \to e\gamma$, $\tau \to \mu\gamma$.

The nice signal's feature of having two  back-to-back high energy
leptons in the final state can be somewhat altered by the
non-monochromaticity of the photon beams produced via Compton
backscattering. So we have restricted the numerical analysis to the
high-energy part of the luminosity spectrum of the photon collider:
on the other hand this part corresponds to collisions of almost
monochromatic and polarized photons, with an integrated luminosity
close to that of the basic lepton collider. Under the very same
conditions we have studied the standard model background and shown
that with suitable cuts it can be taken at the level of
$\sigma_{back} \approx {\cal O} (10^{-2})$ fb. The process
$\gamma\gamma \to ee(\mu\mu)\tau\tau$ presents a configuration with
an undetected $e\tau$ pair emitted at small angle along the
collision axis and with the detected pair of high energy leptons
almost back-to-back, has a potentially large cross section which can
easily mimic the LFV signal. We have considered the signal
statistical significance and found that one can obtain $SS \gtrsim
3$  provided that $\delta_{LL}\gtrsim 10^{-1}$.

\begin{acknowledgments}
M.~C. wishes to thank the ``Fondazione Angelo Della Riccia'' for a
fellowship, the Theory group of
Department of Physics of the
University of Perugia for partial support, and finally
C.~Carimalo and the LPNHE, Universit\`e Pierre et Marie Curie (Paris VI)
for the very kind hospitality.
\end{acknowledgments}

\appendix
\section{Helicity Amplitudes}
\label{helicityamplitudes}

In this appendix we present  explicit expressions for the helicity
amplitudes of the diagrams depicted in Fig.~{\ref{feynman_diagrams}.
\begin{description}
  \item[(a)] \emph{Penguin diagrams} \par These are the diagrams depicted
  in part (a) of Fig.~{\ref{feynman_diagrams}}. We have two types of
  contributions: the chargino-sneutrino loop and the
  slepton-neutralino loop:
  \begin{enumerate}
    \item Chargino-sneutrino
  \begin{equation}\label{penguin_chargino}
    {\cal M}^{\lambda\lambda'}_{i}= +i\frac{e^2(O_{\widetilde{\nu}}^{\widetilde{W}})^2}{16\pi^2}\,
    X^{\lambda\lambda'}_{i} \qquad (i =1,2)
  \end{equation}
  \begin{eqnarray}
  X^{++}_1 &=& +\sin\theta^*\, \left[ (C_0 M_{\widetilde{W}}^2 -2 C_{00})
  + s (C_1+ C_{11}+C_{12})\right] \nonumber \\
  X^{+-}_1 &=& +\sin\theta^*\,  \left(C_0 M_{\widetilde{W}}^2 -2 C_{00}\right)
   \left[\frac{1-\cos\theta^*}{1+\cos\theta^*}\right]\nonumber \\
  X^{-+}_1 &=& -\sin\theta^*\,  \left(C_0 M_{\widetilde{W}}^2 -2 C_{00}\right)\nonumber\\
  X^{--}_1 &=& +\sin\theta^*\,  \left(C_0 M_{\widetilde{W}}^2 -2 C_{00}\right)\nonumber
  \end{eqnarray}
  \begin{eqnarray}
  X^{++}_2 &=& +\sin\theta^*\,  (C_0 M_{\widetilde{W}}^2 -2 C_{00}) \nonumber \\
  X^{+-}_2 &=& +\sin\theta^*\,  \left(C_0 M_{\widetilde{W}}^2 -2 C_{00}\right)
   \left[\frac{1-\cos\theta^*}{1+\cos\theta^*}\right]\nonumber \\
  X^{-+}_2 &=& -\sin\theta^*\,  \left(C_0 M_{\widetilde{W}}^2 -2 C_{00}\right)\nonumber\\
  X^{--}_2 &=& +\sin\theta^*\,  \left[\left(C_0 M_{\widetilde{W}}^2 -2 C_{00}\right)
  + s (C_2+ C_{12}+C_{22})\right] \nonumber
  \end{eqnarray}
  In the above expressions the three-point form factors ($C's$) are
  to be evaluated with the following arguments:
  \[ C_{...} = C_{...}
  (t,0,0,m_{\widetilde{\chi}_j^0}^2,m_{\widetilde{\ell}}^2,m_{\widetilde{\ell}}^2)
  \quad (i=1)
  \]
  \[ C_{...} = C_{...}
  (0,0,t,m_{\widetilde{\chi}_j^0}^2,m_{\widetilde{\ell}}^2,m_{\widetilde{\ell}}^2)
  \quad (i=2)
  \]

    \item Slepton-neutralino
\begin{equation}\label{penguin_slepton}
    {\cal M}^{\lambda\lambda'}_{i}= +i\frac{e^2}{8\pi^2}\,
    \sum_j (O^{\widetilde{\ell}}_{\widetilde{\chi}_j^0})^2
    X^{\lambda\lambda'}_{i} \qquad (i =1,2)
  \end{equation}
  \begin{eqnarray}
  X^{++}_1 &=& +\sin\theta^*\, \left[ +C_{00}  +\frac{s}{4} (1-\cos\theta^*)(C_1+C_{11}+C_{12})
  \right] \nonumber \\
  X^{+-}_1 &=& +\sin\theta^*\,  \left\{ + C_{00}
   \left[\frac{1-\cos\theta^*}{1+\cos\theta^*}\right]
   -\frac{s}{4} (1-\cos\theta^*)(C_1+C_{11}+C_{12})\right\}\nonumber \\
  X^{-+}_1 &=& +\sin\theta^*\,  \left[- C_{00}
  -\frac{s}{4} (1+\cos\theta^*)(C_1+C_{11}+C_{12})\right] \nonumber\\
  X^{--}_1 &=& +\sin\theta^*\,  \left[ +C_{00} -\frac{s}{4} (1+\cos\theta^*)(C_1+C_{11}+C_{12})\right]\nonumber
  \end{eqnarray}
  \begin{eqnarray}
  X^{++}_1 &=& +\sin\theta^*\, \left[ +C_{00}  -\frac{s}{4} (1+\cos\theta^*)(C_2+C_{12}+C_{22})
  \right] \nonumber \\
  X^{+-}_1 &=& +\sin\theta^*\,  \left\{ + C_{00}
   \left[\frac{1-\cos\theta^*}{1+\cos\theta^*}\right]
   -\frac{s}{4} (1-\cos\theta^*)(C_2+C_{12}+C_{22})\right\}\nonumber \\
  X^{-+}_1 &=& +\sin\theta^*\,  \left[- C_{00}
  +\frac{s}{4} (1+\cos\theta^*)(C_2+C_{12}+C_{22})\right] \nonumber\\
  X^{--}_1 &=& +\sin\theta^*\,  \left[ +C_{00} +\frac{s}{4}
  (1-\cos\theta^*)(C_2+C_{12}+C_{22})\right]\nonumber
  \end{eqnarray}
In the above expressions the three-point form factors ($C's$) are
  to be evaluated with the following arguments:
  \[ C_{...}= C_{...}
  (t,0,0,m_{\widetilde{\nu}}^2,M_{\widetilde{W}}^2,M_{\widetilde{W}}^2)
  \quad (i=1)
  \]
\[ C_{...}= C_{...}
  (0,0,t,m_{\widetilde{\nu}}^2,M_{\widetilde{W}}^2,M_{\widetilde{W}}^2)
  \quad (i=2)
  \]
  \end{enumerate}
  \item[(b)] \emph{Self-energy diagrams}
  \begin{enumerate}
    \item external leg corrections
\begin{itemize}
  \item slepton-neutralino
\begin{equation}
      ({\cal M}_{3a}+{\cal M}_{4a})^{\lambda\lambda'} = -i \frac{e^2}{
      16\pi^2}\sum_j O_{\widetilde{\chi}_j^0}^{\ell'}
      O_{\widetilde{\chi}_j^0}^{\ell}\,
      2\, \left[B_0 +B_1\right](0,
      m_{\widetilde{\ell}}^2,m_{\widetilde{\chi}^0_j}^2)\, X^{\lambda\lambda'}\,
      \end{equation}
  \item chargino-sneutrino
  \begin{equation}
    ({\cal M}_{3b}+{\cal M}_{4b})^{\lambda\lambda'} = -i \frac{e^2}{
      16\pi^2}
      (O_{\widetilde{\nu}}^{\widetilde{W}})^2 \,2\,
       \left[B_0 +B_1\right](0,
      m_{\widetilde{\nu}}^2,m_{\widetilde{W}}^2) X^{\lambda\lambda'}\,
  \end{equation}

\end{itemize}
    \item t-channel correction ${\cal M}_5 = {\cal M}_{5a}+ {\cal M}_{5b} $
    \begin{itemize}
    \item slepton-neutralino
      \begin{equation}
      {\cal M}_{5a}^{\lambda\lambda'} = -i \frac{e^2}{
      16\pi^2}\sum_j O_{\widetilde{\chi}_j^0}^{\ell'} O_{\widetilde{\chi}_j^0}^{\ell}
       \left[B_0(t,
      m_{\widetilde{\ell}}^2,m_{\widetilde{\chi}^0_j}^2) +B_1(t,
      m_{\widetilde{\ell}}^2,m_{\widetilde{\chi}^0_j}^2)\right] X^{\lambda\lambda'}\,
      \end{equation}
      \item chargino-sneutrino
      \begin{equation}
      {\cal M}_{5b}^{\lambda\lambda'} = -i \frac{e^2}{
      16\pi^2}
      (O_{\widetilde{\nu}}^{\widetilde{W}})^2 \,
       \left[B_0(t,
      m_{\widetilde{\nu}}^2,m_{\widetilde{W}}^2) +B_1(t,
      m_{\widetilde{\nu}}^2,m_{\widetilde{W}}^2)\right] X^{\lambda\lambda'}\,
      \end{equation}
    \end{itemize}
    The helicity factor is the same in this case:
    \begin{equation}
    X^{\lambda\lambda'} = \frac{\sin\theta^*}{1+\cos\theta^*}\, \left[ \frac{1+\lambda\lambda' +\lambda-\lambda'}{2}
    +\lambda\lambda'\cos\theta^*\right]\, \nonumber
    \end{equation}
    \begin{eqnarray}
        X^{++} &=& +\sin\theta^* \nonumber\\
        X^{+-} &=& +\sin\theta^*\frac{1-\cos\theta^*}{1+\cos\theta*} \nonumber\\
        X^{-+} &=& -\sin\theta^* \nonumber\\
        X^{--} &=& +\sin\theta^* \nonumber
      \end{eqnarray}
  \end{enumerate}
  \item[(c)] \emph{Sea-Gull and Box diagrams}\par These are the diagrams depicted
  in part (c) of Fig.~\ref{feynman_diagrams}.
\begin{itemize}
\item (Seagull) The seagull type
  diagram has only  a contribution from a slepton-neutralino loop (${\cal M}_6$).
\begin{equation}{\cal M}^{\lambda\lambda'}_6=0 \nonumber \end{equation}
\item (Scalar Box) \par This is the slepton neutralino box diagram in
part(c) of Fig.~\ref{feynman_diagrams} (${\cal M}_7$).
\begin{eqnarray}
   {\cal M}^{\lambda\lambda'}_7&=& +i \frac{e^2}{16\pi^2}
   \sum_j (O_{\widetilde{\chi}_j^0}^{\widetilde{\ell}})^2\,
   4 \left(\frac{s}{2}\sin\theta^*\right)\, \left[
   \frac{1+\lambda\lambda'}{2}D_{002}\right.\nonumber \\
   &&-\left(\frac{\lambda-\lambda'}{2}+\lambda\lambda'\cos\theta^*\right)
   (D_{00}+D_{001}+D_{002}+D_{003}) \nonumber \\
   && -\frac{s}{8}\lambda\lambda'\sin^2\theta^*\left(D_{112}
   +2D_{122}+2D_{123}+D_{222}+2D_{223}+D_{233} \right. \nonumber \\
   &&\left. \left. +D_2 +2 ( D_{12}+D_{22}D_{23} \right)\phantom{\frac{1}{2}}\right]
\end{eqnarray}
The $D$ form factors appearing in the above formula for the scalar
box diagram are to be evaluated with the following arguments:
\[
D_{...} = D_{...}
(0,0,0,0,t,s,m_{\chi_j^0}^2,m_{\widetilde{\ell}}^2,m_{\widetilde{\ell}}^2,m_{\widetilde{\ell}}^2)
\]
\item Chargino-sneutrino loop \par This is the box diagram involving fermions
depicted as (${\cal M}_8$) in part(c) of
Fig.~\ref{feynman_diagrams}. \begin{equation} {\cal
M}_8^{\lambda\lambda'} = -i \frac{e^2}{16\pi^2}\,
(O_{\widetilde{\nu}}^{\widetilde{W}})^2\,  \left[ \sum_{i} \langle i
\rangle^{\lambda\lambda'} D_i+ \sum_{ij} \langle
ij\rangle^{\lambda\lambda'} D_{ij}+\sum_{ijk} \langle
ijk\rangle^{\lambda\lambda'} D_{ijk}\right]
\end{equation}
The various (non-zero) coefficients multilplying the four-point loop
form factors ($D_i, D_{ij}, D_{ijk}$)  are given below:
\begin{eqnarray}
\langle 0 \rangle^{\lambda\lambda'}&=&-M_{\widetilde{W}}^2\,
\frac{s}{4}\, \sin\theta^*\left(1+\lambda\lambda' +\lambda -\lambda'
+2\lambda\lambda'\cos\theta^*\right)\nonumber\\
\langle 1 \rangle^{\lambda\lambda'}&=&+M_{\widetilde{W}}^2\,
\frac{s}{2}\, \lambda' \sin\theta^*\left(1 -\lambda\cos\theta^*\right)\nonumber\\
\langle 2 \rangle^{\lambda\lambda'}&=&+M_{\widetilde{W}}^2\,
\frac{s}{4}\, \sin\theta^*\left(1+\lambda\lambda' -(\lambda
-\lambda')
-2\lambda\lambda'\cos\theta^*\right)\nonumber\\
\langle 3 \rangle^{\lambda\lambda'}&=&-M_{\widetilde{W}}^2\,
\frac{s}{2}\, \lambda \sin\theta^*\left(1 +\lambda'\cos\theta^*\right)\nonumber\\
\langle 00
\rangle^{\lambda\lambda'}&=&+\frac{s}{2}\,\sin\theta^*\left[
1+\lambda\lambda' +(\lambda -\lambda')
+2\lambda\lambda'\cos\theta^*\right]\nonumber\\
\langle 12
\rangle^{\lambda\lambda'}&=&+\frac{s^2}{4}\,\sin\theta^*\,
\lambda'(1-\lambda)(1+\cos\theta^*)\nonumber\\
\langle 13
\rangle^{\lambda\lambda'}&=&-\frac{s^2}{4}\,\sin\theta^*\left[
1+\lambda\lambda' +(\lambda -\lambda')
+2\lambda\lambda'\cos\theta^*\right]\nonumber\\
\langle 22 \rangle^{\lambda\lambda'}&=&-\frac{s^2}{8}
\,(\lambda-\lambda')(1-\lambda)\,\sin\theta^*(1+\cos\theta^*)\nonumber\\
\langle 23 \rangle^{\lambda\lambda'}&=&-\frac{s^2}{4}\,
\lambda(1+\lambda')\,\sin\theta^*\,(1+\cos\theta^*)\nonumber
\end{eqnarray}
\begin{eqnarray}
\langle 001
\rangle^{\lambda\lambda'}&=&-s\sin\theta^*[\lambda+\lambda' +
\lambda'(1-\lambda\cos\theta^*)]\nonumber \\
\langle 002\rangle^{\lambda\lambda'}&=& (\frac{s}{2}\sin\theta^*)
[-(1+\lambda\lambda') +\lambda(1+\lambda'\cos\theta^*)-\lambda' (1-\lambda\cos\theta^*) ]
\nonumber \\
\langle 003\rangle^{\lambda\lambda'}&=& +{s}\sin\theta^*\,
[\lambda(1+\lambda'\cos\theta^*)+\lambda+\lambda']\nonumber \\
\langle 112\rangle^{\lambda\lambda'}&=&+\frac{s^2}{4}\,
\lambda'(1-\lambda)\sin\theta^*(1+\cos\theta^*)\nonumber\\
\langle 113\rangle^{\lambda\lambda'}&=&+\frac{s^2}{2}\,
\lambda'\sin\theta^*(1-\lambda\cos\theta^*)\nonumber\\
\langle 133 \rangle^{\lambda\lambda'}&=&-\frac{s^2}{2}\,
\lambda\sin\theta^*(1+\lambda'\cos\theta^*)\nonumber\\
\langle 122 \rangle^{\lambda\lambda'}&=&-\frac{s^2}{8}\,
(3\lambda'-\lambda)(1-\lambda)
\sin\theta^*(1+\cos\theta^*)\nonumber\\
\langle 123\rangle^{\lambda\lambda'}&=&+\frac{s^2}{2}\sin\theta^*\,
\left[\lambda'(1-\lambda\cos\theta^*)
-\frac{1}{2}(1+\lambda\lambda'+\lambda+\lambda')\right]\nonumber\\
\langle 132\rangle^{\lambda\lambda'}&=&\delta_{\lambda+}\,
\frac{s^2}{2}\,\sin\theta^*\,(1+\cos\theta^*)\nonumber\\
\langle 222 \rangle^{\lambda\lambda'}&=&-\frac{s^2}{8}\,
(\lambda-\lambda')(1-\lambda)\sin\theta^*
(1+\cos\theta^*)\nonumber\\
 \langle 223 \rangle^{\lambda\lambda'}&=&
-\frac{s^2}{8}\lambda(1+\lambda')(3-\lambda)\,
\sin\theta^*(1+\cos\theta^*)\nonumber\\
 \langle 233\rangle^{\lambda\lambda'}&=&-\frac{s^2}{4}\,
\lambda(1+\lambda')\,\sin\theta^*\,(1+\cos\theta^*)\nonumber
\end{eqnarray}
\end{itemize}
The $D$ form factors appearing in the above formula for the chargino
sneutrino box diagram are to be evaluated with the following
arguments:
\[
D_{...} = D_{...}
(0,0,0,0,t,s,m_{\widetilde{\nu}}^2,m_{\widetilde{W}}^2,m_{\widetilde{W}}^2,m_{\widetilde{W}}^2)
\]
\end{description}

A diagram with a LFV and a LFC scalar line, for example, is described
by the propagators of Eq.~(\ref{LFVprop}) and Eq.~(\ref{LFCprop}), so that the loop coefficients
in the amplitudes are a sum of four integrals, while in the diagrams with only
a LFV line, are a sum of two.
The scalar two point function $B_0$ and the tensor coefficients
$B_1$, $C_{00}$ that appear in
the electroweak penguins are ultra-violet divergent, but the amplitudes are
finite due the ortogonality of the slepton mixing matrix.
The rule for obtaining the helicity
amplitudes for the exchanged diagrams from those of the direct
diagrams:
\begin{equation}
{\cal M}^{\lambda\lambda'}_{exch.}(\sin\theta^*,\cos\theta^*) ={\cal
M}^{\lambda'\lambda}_{direct}(-\sin\theta^*,-\cos\theta^*)
\end{equation}
The loop form factors are exchanged accordingly to the same rule:
($\cos\theta^* \to -\cos\theta^*$, and $\sin\theta^* \to
-\sin\theta^*$).

\end{document}